\documentstyle[aps,prl,multicol,rotate,psfig]{revtex}

\newcommand{\be}{\begin{equation}}
\newcommand{\ee}{\end{equation}}
\newcommand{\ba}{\begin{eqnarray}}
\newcommand{\ea}{\end{eqnarray}}
\newcommand{\bt}{\begin{tabular}}
\newcommand{\et}{\end{tabular}}
\newcommand{\bc}{\begin{center}}
\newcommand{\ec}{\end{center}}
\newcommand{\ben}{\begin{enumerate}}
\newcommand{\een}{\end{enumerate}}
\newcommand{\bi}{\begin{itemize}}
\newcommand{\ei}{\end{itemize}}
\newcommand{\bmpage}{\begin{minipage}}
\newcommand{\empage}{\end{minipage}}

\def \df {{\rm d}}

\def\la {\lambda}

\begin{document}
\draft

\title{Cumulant ratios and their scaling functions \\
       for  Ising systems in a strip geometry}
\author{$ ^1$Andrzej Drzewi\'nski, $ ^2$Jacek Wojtkiewicz}
\address{$ ^1$Institute of Low Temperature and Structure Research, 
Polish Academy of Sciences,\\
 P.O.Box 1410, 50-950 Wroc\l aw 2, Poland.}
\address{$ ^2$Department for Mathematical Methods in Physics, \\
Warsaw University, 00-682 Warszawa, Poland.}
\date{\today}
\maketitle

\begin{abstract}
We calculate the fourth-order cumulant ratio (proposed by
Binder) for the two-dimensional Ising model in a strip geometry
$L\times\infty$. The Density Matrix Renormalization Group
method enables us to consider typical open boundary conditions
up to $L=200$. Universal scaling functions of the cumulant ratio
are determined for strips with parallel as well as opposing
surface fields.
\end{abstract}


\begin{multicols}{2} \narrowtext

{\bf Introduction.}
An {\em universality principle} is a cornerstone of contemporary
theory of phase transitions. According to this principle, the
following sorts of quantities are  universal: critical
exponents, certain amplitude ratios and scaling functions
\cite{PHA91}. They differ each from other in their status.  The
(bulk) critical exponents are independent on boundary conditions,
 whereas two other groups
{\em are} dependent.

The critical exponents are known for many models (both exactly
and approximately).  The collection of results available  for
amplitude ratios is also rich, but significantly smaller than
for exponents;   see in Ref. \cite{PHA91} for  exhaustive
information.

Among amplitude ratios, so called {\em cumulant ratios} are of
great importance. They supply some information on scaling
functions (cumulants are proportional to derivatives of these
functions at zero values of argument(s)); they measure deviation
of magnetization fluctuations at criticality from gaussian
distribution; moreover, they are closely related to some versions
of renormalization group  \cite{Binder81} (it gains also reflection
in terminology: cumulant ratios are customarily termed ``renormalized
coupling  constant'' in field theory). Cumulant ratios have also been
used to locate the critical points and critical
 lines in many models \cite{Binder81,Kam96,BLF,SalDer85}.

Most results for cumulants have been obtained for (partially or
completely) periodic boundary conditions. For most extensively studied
the two-dimensional Ising model, numerous  results are available 
\cite{Binder81,SalDer85,Bruce81,Bruce85,BurDer85,Desai88,FishZin,KamBlo93,SaSo,SokOrl}.
However, there are also other, very natural "open" boundary conditions:
``free'' (no surface fields), ``wall$++$'' (infinite
parallel surface fields), ``wall$+-$'' (infinite opposing surface fields).
For these open boundary conditions, the number of results is very
small. Only  papers we know where such results are available 
are \cite{Binder81,BLF}. Motivated by this situation, we
state  the aim of this paper:

{\em Calculation of universal cumulant ratios for the two-dimensional
 Ising model in a strip
geometry under the following boundary conditions: ``free'', ``wall$++$'',
 ``wall$+-$''.}

We have calculated cumulant ratios using method called the Density Matrix
Renormalization Group (DMRG). Since the DMRG is most powerful
for open boundary conditions, it is particularly suited for our goals.

{\bf Definition of cumulants.}
We consider the two-dimensional Ising system on a square lattice
 in a strip geometry 
($L$ is width of the strip
and  $N$ is its length)
with the Hamiltonian

\be
{ {\cal H} } = -J\left[
\sum_{<i,j>}s_i s_j - H \sum_i s_i 
-H_1 \sum_i^{(1)}s_i
-H_L\sum_i^{(L)}s_i
\right],
\label{Ham} 
\ee
where the first sum runs over all nearest-neighbour pairs of sites
while the last two sums run over the first and the $L$-th column,
respectively. $H$ is the bulk magnetic field, whereas $H_1$ and $H_L$
are the surface fields. $H$, $H_1$ and $H_L$ are  dimensionless quantities
(all of them are measured in units of $J$).
 In the course of calculations of the 
cumulant
in the termodynamic limit, two limiting processes are taken: $T \to T_c$
and $L \to \infty$. In general, a value of the cumulant do depend on ordering
of these limits \cite{SaSo}. In our paper 
 we analyze so-called "massless" case 
  (analogously as  in \cite{SaSo,KimPa}):
  $T=T_c= 
2\slash\ln(1+\sqrt{2})$
$\simeq 2.269185$ followed by $L \to \infty$.
Therefore, we do not notice the
temperature dependence below. 
We also drop (as unnecessary) explicit
dependence on surface fields until discussion of scaling
functions.

We consider  the ratio of moments of magnetization proposed by Binder 
\cite{Binder81}.
Definitions of cumulant ratios for a system in a strip geometry 
have been widely presented in the literature 
\cite{Binder81,SalDer85,BurDer85,KamBlo93}.
Let us first define
\be U_L =
\lim_{N\to\infty}[N(1-\frac{1}{3}<M^4><M^2>^{-2})],
\ee
where $M = \sum_i s_i$ is the total (extensive) magnetization. Then, the 
cumulant ratio 
$A_U$ in question is 
\be A_U = \lim_{L\to\infty}L^{-1} U_L.
\ee

An equivalent (but more convenient for us)  formula for the above cumulant
is as follows: Let $\la(L;H)$ be the largest eigenvalue of transfer matrix
for the strip of width $L$ ($-T\log\la(L;H)$ is the free energy for one column 
of spins). 
Define
\be m_2(L) =
\frac{\df^2}{\df H^2}\log\la(L;H)\vert_{H=0},
\label{der2}
\ee
\be
m_4(L) = \frac{\df^4}{\df H^4}\log\la(L;H)\vert_{H=0}.
\label{der4}
\ee 
Then our cumulant is equal to\cite{SalDer85}
\be
A_U = \lim_{L\to\infty}r(L) \equiv 
 \lim_{L\to\infty}m_4(L)\slash 3 L m_2^{2}(L).
\label{main}
\ee

Our method of calculation is  based directly on the definition (\ref{main}). 
We  first find   logarithms of the largest eigenvalue
 $\la(L;H)$ for some values of $H$ (at fixed  $L$).
 Next, we calculate numerically 
derivatives  (\ref{der2}) and (\ref{der4}), then the ratio $r(L)$,
 and finally perform the extrapolation $L\to\infty$.

{\bf Some technical details of calculations}. We use the
DMRG method  for calculations of
$\log\la(L;H)$.  Originally, this method has been proposed by
 White  \cite{White}
for finding accurate approximations to the ground state and 
the low-lying excited states of quantum chains. Its heart 
 is recursive construction of the effective Hamiltonian of a very large
 system using a 
truncated basis set, starting from an exact solution for small systems.
Later, the DMRG was adapted by  Nishino \cite{Nis95} for
two-dimensional classical systems.
 The   DMRG  has been applied successfully
to many different problems and now it can be treated as a 
standard method, which is
 very flexible, relatively easy to implement and very precise.
 For a comprehensive review 
of a background, achievements and limitations of DMRG, see in Refs. 
 \cite{DMRG,KH}.

A factor crucial for precision of DMRG is so-called {\em number of 
states kept} $m$, describing the dimensionality of effective 
transfer matrix \cite{White,DMRG}; the  larger number
of states kept, the more accurate the value of the free energy.
Using $m=50$ we can calculate the free energy with accuracy of the
order $10^{-12}$ for strips of
width of the order $L=200$.  This is an one order
more than size of systems available by exact diagonalization of
transfer matrix. This  fact is crucial for us, because of using the
extrapolation procedures. In our calculations we apply the
 {\em finite system algorithm}, 
developed by White for studying finite  systems \cite{White}.  
Additional factor determining the accuracy
 of the method is {\em number of sweeps}, i.e. number
of iterations made in order to obtain self-consistency of results.
 Our numerical experience shows that in most cases, it
is sufficient to apply only one sweep (although in the "wall$+-$" case
 two sweeps are necessary -- see below).

In our calculations of cumulant ratios, we have also a factor
limiting accuracy 
that is independent on the DMRG method: {\em
 accuracy of numerical differentiation.}
  In the procedure of
 numerical differentiation, a suitable choice of increment $\Delta H$ of an
 argument is of crucial importance.
It is clear that $\Delta H$ should be taken as small as possible;
on the other hand, due to finite accuracy in calculation of $\lambda$,
an error of difference quotient increases with decreasing $\Delta H$.
The increments used in
our calculations have been determined as a compromise between above
two tendencies.  Additional factor determining the accuracy of numerical
differentiation, is a number of points used to calculate the
derivative.  We use formulas where a derivative is determined from the
second-order Taylor expansion (i.e. we need $n+3$ values of function
for $n$-th derivative; this way, an accuracy is of the order
${\cal{O} }((\Delta H)^3)$.)  
Therefore, the $m_2$ was
determined from $5$ points ($3$ points in symmetrical case, 
i.e. $f(H) = f(-H)$) and from $7$ points for $m_4$  ($4$ points when
the symmetry was present).

We have tested correctness of our calculations in several ways.  One
of them was $L$-dependence on derivatives   $m_2$ and $m_4$.  Finite-Size
 Scaling (FSS) theory \cite{PHA91,Barber83}   predicts the
following dependence on the $n$-th derivative of the
 free energy as a function of the system size  $L$: 
\be
 \frac {\df^n f}{\df H^{n}}(L)\vert_{H=0} \sim
L^{-\tilde{d}+ n \Delta\slash \nu},
 \ee 
where we have $\Delta=15\slash 8$ and $\nu=1$ for the two-dimensional Ising 
model.
$\tilde{d}$ is a
dimension of system in  ``finite-size direction'', i.e. it is a number
of linearly independent directions along which a size of the system is
finite. 
For finite  systems (for instance a torus), 
 $\tilde{d}$  is equal to space dimension of the
system. In our case,
 the system is
infinite in one direction (along the strip) and finite in the second
direction (across the strip), so we have to take  $\tilde{d}=1$. This
assumption gives the following predictions for derivatives:  
\be m_2
\sim L^{\rho_2}, \;\;\;\;\; m_4 \sim L^{\rho_4},
\label{expro2iro4}
 \ee 
where $\rho_2 =
{11\slash 4}$ and $\rho_4 = {13\slash 2}$.

An extrapolation procedure has been performed with use of the powerful
BST method \cite{HenSch88}.

{\bf Results: the ``free'' case. } It corresponds to zero surface fields
$H_1 = H_L =0$
in the formula (\ref{Ham}). 
We have performed calculations for $L$ in the range 
 $160 \leq L \leq 200$
 with  step
 $10$; these values of $L$ were taken in all situations. 
  We took an increment of ``bulk'' magnetic field
 $\Delta H =5\times 10^{-6}$, $m=50$ and one sweep.
  The results are listed in the Table.

As a byproduct, we have tested the FSS predictions for $L$-dependence
of derivatives   $m_2$ and $m_4$.
Values of corresponding exponents (see Eq. (\ref{expro2iro4}) ) are:
 $\rho_2 = 2.7495(3)$ and $\rho_4 = 6.50(3)$, so
predictions of FSS are confirmed in excellent manner.
The same conclusion is true in next two situations.

As another test of correctness (and quality) of DMRG results,
we have calculated ratios by immediate
{\em numerical diagonalization of transfer
matrix} for $10\leq L\leq 18$ (L even;
these values of L have also been used in the next  cases).
 We proceeded as above, i.e. by calculation
of logarithm of the largest eigenvalue for some values of bulk field $H$,
followed by numerical differentiation of $f(H)$ and computation 
of ratio and extrapolation, without any ``renormalization''.
 We took an increment  $\Delta H = 10^{-4}$. 
 We have obtained   
 $A_U= -1.094(1)$; $\rho_2 = 2.746(1)$; $\rho_4 = 6.46(1)$. It is seen that 
the results are fully consistent with the  DMRG calculations
but less precise;
 we have the same situation for two other boundary conditions.

{\bf Results: the ``wall$++$'' case. }
The ``wall$++$'' boundary condition 
corresponds to the assumption that all boundary spins
have the same value and sign.  
It is equivalent to putting $H_1 = H_L = \infty$ in
(\ref{Ham}). Numerical
experience suggests that it is sufficient to take  $H_1 = 10 $ -- for
larger values of $H_1$ the changes of the free energy are negligible
\cite{CDR}.

The ``wall$++$'' configuration
 is more intricate, from numerical point
of view, than ``free'' system. The complication is due to the fact
that, for parallel surface fields of the same sign,
 the maximum of the free energy $f(H)$
does not appear for $H=0$ but it is shifted to a certain non-zero value
$H_0(L)$. This phenomenon  is
called the {\em capillary condensation}  \cite{CDR,FishNak}.
In order to calculate derivatives and ratios at zero
magnetization (i.e. at the maximum of the free energy),
we  first have to find its position $H_0(L)$.  FSS
predicts \cite{FishNak} the following dependence: $H_0(L)\sim
 L^{-\Delta\slash \nu}$.
 From our DMRG calculations, we have obtained
the  value  $\Delta\slash\nu= 1.8749(2)$.

For the ``wall$++$'' configuration the free energy is not longer a
 symmetric function
of the bulk field $H$, so we have been forced to calculate $m_2$ from 5
points and $m_4$ from 7 points. We have taken an increment  
$\Delta H = 5\times
10^{-6}$, $m=50$ and one sweep.
 The results are presented in the Table.  For
exponents of $m_2$ and $m_4$, we have obtained:
 $\rho_2 = 2.7504(3)$ and $\rho_4 =
6.5024(3)$.
The precision of these results is a little bit less than in
the ``free'' case, although still very satisfactory (three significant
digits).
 However, it should be stressed that here we must do
much more numerical computations than in the ``free'' case, so some lack
of precision is inevitable.
 
Exact
diagonalization of transfer matrix 
 gave the following
values: $A_U= 0.45(4)$; $\rho_2 = 2.75(1)$; $\rho_4 = 6.5(2)$.  

{\bf Results: the ``wall$+-$'' case.}  One of important physical
 implications of
``$+-$'' boundary condition is the presence of
an {\em interface} between ``$+$'' and ``$-$'' phases in the system. It
causes large fluctuations, which have an implication
 in numerical practice,
 namely, {\em two sweeps} are necessary
 to ensure self-consistency of results. In our calculations, the
value of surface field $H_1=100$ 
 an 
increment $ \Delta H =2\cdot 10^{-5}$, and $m=40$ were taken.
The results are listed in the Table. Values of exponents 
 are: $\rho_2=2.7502(2)$, $\rho_4=6.502(2)$.

The procedure of exact diagonalization of transfer matrix
gave the following
values: $A_U=-0.305(2)$; $\rho_2 =2.755(1)$; $\rho_4 =6.50(2)$.

As a matter of some interest, 
let us remark that for the ``wall$+-$'' boundary condition
the $L$ dependence is much weaker than for the ``free'' and ``wall$++$'' 
situations.

\vskip.5cm
\centerline{{\bf Table.}\em Values of cumulant ratios for some values of $L$}
\centerline{
\begin{tabular}{|c|c|c|c|}
\hline\hline
$L$ &$r(L)$, free&$r(L)$, wall$++$&$r(L)$, wall$+-$\\
\hline
160 & -1.098525 & 0.462556 &-0.304831\\
170 & -1.098234 & 0.462225 & -0.304859\\
180 & -1.097964 & 0.461859 &-0.304883\\
190 & -1.097723 & 0.461525 &-0.304902\\
200 & -1.097481 & 0.461133 &-0.304915\\
\hline
$\infty$&-1.0932(3)& 0.455(2) &-0.3050(1)\\
\hline\hline
\end{tabular}
}
\vskip.32cm

{\bf Scaling functions for ratios.}
The "wall$++/+-$"-type conditions can be treated as limiting case
of the system with equal {\em finite} parallel/antiparallel ($++/+-$) 
surface fields. Another limiting case is the "free" boundary condition, 
where the values of surface fields are set to zero.
One can expect that for intermediate situations, i.e.
finite values of boundary field $H_1$, 
cumulants would be smooth functions of $H_1$. Particularly interesting
are {\em scaling properties} of these  functions. 
The scaling theory predicts
that at criticality, the system depends only
 on one variable, namely
dimensionless combination $\zeta=L H_1^2$ \cite{FishdeGen}.
In the
other words, we
expect that the cumulant $r$ should depend on the surface field $H_1$
and strip width $L$ only by combination $\zeta$.
 We have calculated $r(L,H_1)$ for both 
 "$++$" and "$+-$" boundary conditions
for $L=40,80,120$ using $m=40$, and at full range of scaling variable.
The results are presented on Figs. 1a and 1b. It is seen that scaling
properties are confirmed in excellent manner.

Limiting values of these functions (i.e. for $\zeta=0$ and $\zeta\to \infty$) 
are fully consistent with our more precise calculations,
although convergence of ratio to its limit value is much faster 
for "wall$+-$" than for "wall$++$".
As far we know,  scaling functions for cumulants have 
been almost not studied so far. Only exception is paper \cite{OK};
however, the authors consider scaling functions different from ours.

{\bf Summary.}
We have calculated cumulant ratios for Ising strips with three
natural boundary conditions,
 almost
  not studied so far: ``free'',
``wall$++$'' and ``wall$+-$'' situations. 
We have applied the Density Matrix 
Renormalization Group method  followed by  numerical differentiation
and extrapolation $L\to\infty$.
 We claim
that our results are very precise (three or four significant digits).
 The precision is comparable with
three other ``top quality'' methods used in similar calculations: Monte
Carlo \cite{SaSo}, some versions of Renormalization Group  \cite{SokOrl}
and
analysis of high-temperature series  \cite{FishZin}. 

\begin{figure}[b]
\centerline{
\rotate[r]{\psfig{file=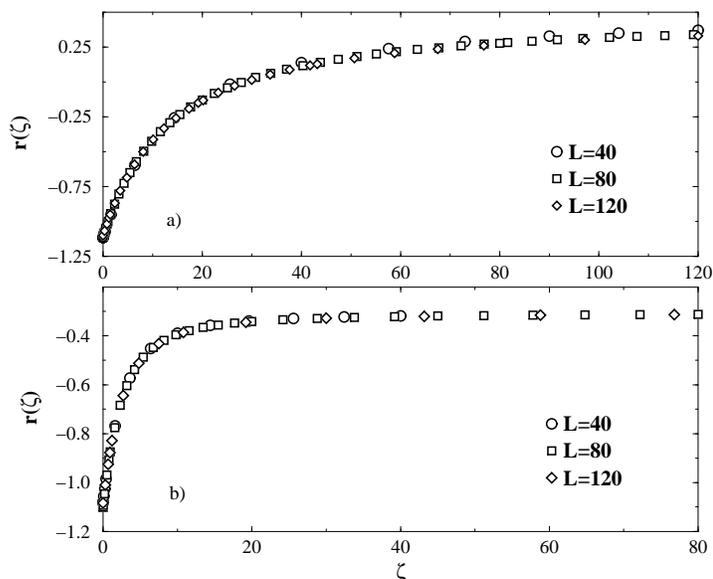,height=9.5cm}}}
\vskip 0.2truecm
\caption{The cumulant ratio as a function of scaling variable
$\zeta=L H_1^2$: {\em a)} the ``$++$'' boundary condition;
{\em b)}  the ``$+-$'' boundary condition. Notice much weaker $L$
dependence for ``$+-$''  than for ``$++$'' situation.
Moreover, the convergence of ratio to its limit value,
when $\zeta\to\infty$ is much faster
for "$+-$" than for "$++$". For the latter case, the saturation
is achieved for $\zeta\approx 500$ (outside the range of plot).}
\label{FIG}
\end{figure}

We have also calculated the quantity which apparently escaped
the attention so far (at least for boundary conditions considered by us),
 namely, the scaling functions for cumulants.
Such functions give information of how finite surface fields influence
values of cumulants. This influence is significant -- in one case ("$++$")
even the sign of cumulant changes upon growth of surface field. We do not
know how fundamental this phenomenon is. At first glance, it seems to be
related to the lack of symmetry (i.e. $f(H_0+H) \neq f(H_0-H)$) and to
a non-zero value of the third derivative of the free energy $f$ with respect to 
$H$
at $H_0$. However, the other explanations are not excluded and we will discuss
it elsewhere.

Natural lines of continuation of our investigations are:
testing of universality of cumulants and scaling functions
(for other models in the two-dimensional Ising universality
class, for example the hard squares model) and 
calculation of higher cumulants. This work is currently in progress. 

Most of our numerical calculations were performed on Pentium II machines
under Linux. We used ARPACK package to calculate eigenvalues.

{\bf Acknowledgments}. This work was partially funded by KBN Grant
No. 2P03B10616.

\end{multicols}
\end{document}